\documentclass[a4paper]{jpconf}
\usepackage{graphicx}
\usepackage{amssymb}
\usepackage{amsmath}
\usepackage{latexsym}
\def\vev#1{\left\langle #1\right\rangle}
\begin{document}
\title{Dark matter stability from non-Abelian discrete flavor symmetries}

\author{E. Peinado}

\address{ AHEP Group, Institut de F\'{\i}sica Corpuscular --
  C.S.I.C./Universitat de Val{\`e}ncia \\
  Edificio Institutos de Paterna, Apt 22085, E--46071 Valencia, Spain}

\ead{epeinado@ific.uv.es}

\begin{abstract}
We present a mechanism for the dark matter stability  in the framework of a non-Abelian flavour symmetry renormalizable model. The same non-abelian discrete flavor symmetry which accounts for the observed pattern of neutrino oscillations, spontaneously breaks to a $Z_2$ subgroup which renders DM stable. The simplest scheme leads to a scalar doublet DM potentially detectable in nuclear recoil experiments, inverse neutrino mass hierarchy, hence a neutrinoless double beta decay rate accessible to upcoming searches.\end{abstract}
\section{Introduction}
Non-baryonic Dark Matter (DM) is one of the most compelling 
problems of modern cosmology.
Despite the fact that its existence is well established by cosmological and astrophysical
probes, its nature remains elusive. 
Still, observations can constraint the properties of dark matter and give some hints about its identity.
For instance, a fundamental
requirement for a viable dark matter candidate is the stability over cosmological times. This suggests the existence of
an exact or slightly-broken symmetry protecting or suppresing its decay.

It has been shown recently that such symmetry can also be related
to the flavor structure of the Standard Model~\cite{arXiv:1007.0871}. The model proposed in ~\cite{arXiv:1007.0871} is based on a $A_4$ symmetry with four $SU(2)$ Higgs doublets. After the electroweak symmetry breaking, the $A_4$ (even permutation of four objects) group is spontaneously broken into a $Z_2$ subgroup which is responsible for the DM stability. The leptonic sector is also extended. It consist in four right handed neutrinos and the light neutrino masses are generated through the type-I seesaw mechanism and obey an inverted hierarchy mass spectrum with $m_{\nu_3}=0$ and vanishing reactor angle $\theta_{13}=0$. 

We have also been consider models with a different matter content for the right handed neutrinos with the same DM stability mechanism but with different neutrino phenomenology~\cite{arXiv:1011.1371} or a model based on the dihedral group $D_4$ where the some flavour changing neutral currents are present and constraint the DM sector~\cite{arXiv:1104.0178}. Other models with flavor symmetries but with decaying DM have also been considered, see for instance~\cite{arXiv:1010.4963,arXiv:1011.5753}. For a model with stable DM but with the same behavior of the DM, in the sense that instead of being inert totally, the DM couples to some right handed neutrinos in a similar way in our model\footnote{This afther the $A_4$ is broken into the $Z_2$ symmetry.} see~\cite{arXiv:1111.4938}. 
\section{The Model and DM stability}
\begin{figure}
\begin{center} 
\includegraphics[width=0.3\textwidth]{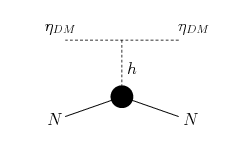}
\includegraphics[width=0.3\textwidth]{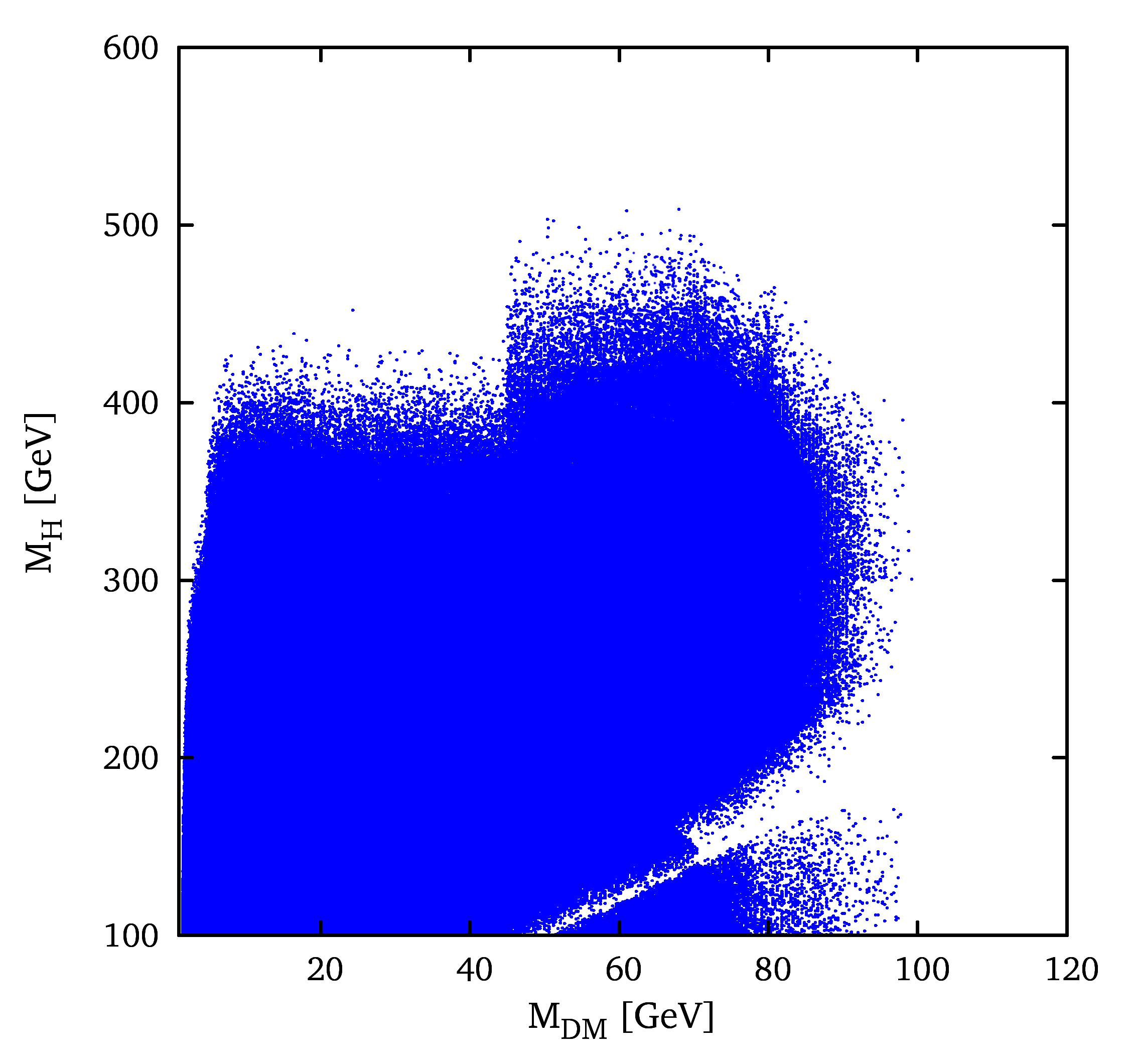}
\includegraphics[width=0.3\textwidth]{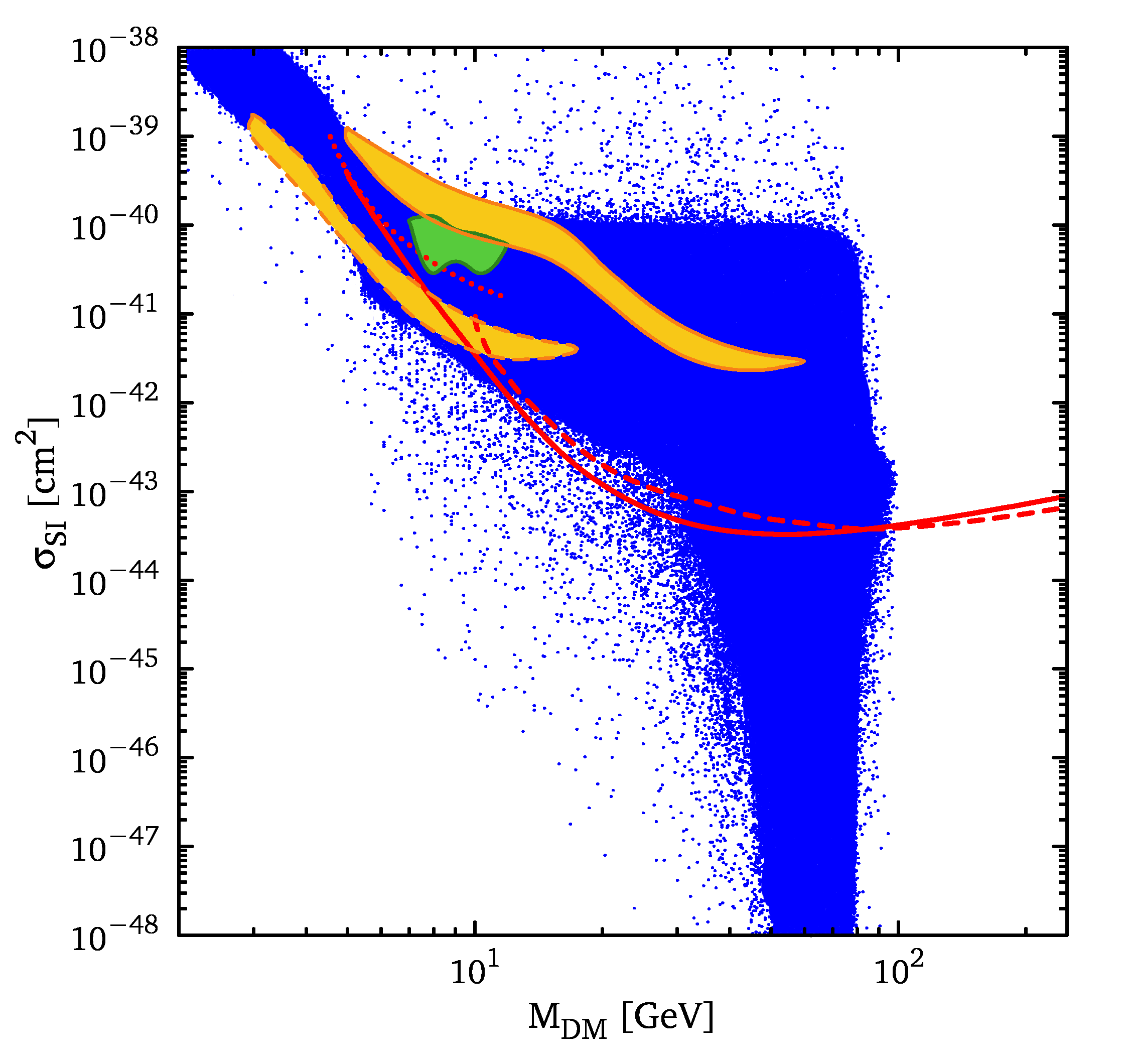}
\caption{Feynman diagrams for elastic
  scattering (left). Regions in the plane DM mass (MDM) - lightest Higgs boson MH allowed by collider constraints and leading to a DM relic abundance compatible with WMAP measurements (center). Spin-independent DM scattering cross section ofoff-protons as a function of the dark matter mass. The orange
regions delimited by the dashed (solid) line show the DAMA/LIBRA annual modulation regions including (neglecting) the
channeling effect. The green region corresponds to the COGENT data. Dashed and dotted red lines correspond to the upper bound from CDMS. XENON100 bounds are shown as a solid red line. }
\label{diagram}
\end{center}
\end{figure}

The model consist on the SM matter fields plus three Higgs doublets which transform as a triplet under $A_4$, and four right handed neutrinos transforming as a singlet and a triplet of the flavor group. In table~\ref{tab1} we present the relevant quantum numbers for the matter fields.
\begin{table}[h!]
\begin{center}
\begin{tabular}{|c|c|c|c|c|c|c|c|c||c|c|}
\hline
&$\,L_e\,$&$\,L_{\mu}\,$&$\,L_{\tau}\,$&$\,\,l_{e}^c\,\,$&$\,\,l_{{\mu}}^c\,\,$&$\,\,l_{{\tau}}^c\,\,$&$N_{T}\,$&$\,N_4\,$&$\,H\,$&$\,\eta\,$\\
\hline
$SU(2)$&2&2&2&1&1&1&1&1&2&2\\
\hline
$A_4$ &$1$ &$1^\prime$&$1^{\prime \prime}$&$1$&$1^{\prime \prime}$&$1^\prime$&$3$ &$1$ &$1$&$3$\\
\hline
\end{tabular}\caption{Summary of relevant model quantum numbers}\label{tab1}
\end{center}
\end{table}
The group of even permutations of four objects, $A_4$, has two generators: $S$, and $T$, which obeys the properties $S^2=T^3=(ST)^3=I$. $S$ is a $Z_2$ generator while $T$ is a $Z_3$ generator. $A_4$ has four irreducible representations, three singlets $1$, $1^\prime$, and $1^{\prime \prime}$ and one triplet. The generators of $A_4$ for each irreducible representation are
\begin{equation}\begin{array}{lr}
\begin{array}{ccc}
1 & S=1 & T=1\\
1^\prime & S=1 & T=\omega^2\\
1^{\prime\prime} & S=1 & T=\omega
\end{array}& 3:\begin{array}{lr}
S=\left(
\begin{array}{ccc}
1&0&0\\
0&-1&0\\
0&0&-1\\
\end{array}
\right)&T=\left(
\begin{array}{ccc}
0&1&0\\
0&0&1\\
1&0&0\\
\end{array}
\right)\end{array}
\end{array}\label{generators}
\end{equation}
where $\omega^3=1$. 
The Yukawa Lagrangian of the model is 
\begin{eqnarray}\label{lag}
\mathcal{L}&=&y_e L_el_{_e}^c H+y_\mu L_\mu l_{_\mu}^c H+y_\tau L_\tau l_{_\tau}^c H++y_1^\nu L_e(N_T\eta)_{1}+y_2^\nu L_\mu(N_T\eta)_{1''}+\nonumber\\&&+y_3^\nu L_\tau(N_T\eta)_{1'}+y_4^\nu L_e N_4 H+ M_1 N_TN_T+M_2 N_4N_4+
\mbox{h.c.}
\end{eqnarray}
This way $H$ is responsible for quark and charged lepton masses, the latter automatically diagonal\footnote{For quark mixing angles generated through higher dimension operators see reference ~\cite{arXiv:1104.5676}.}. Neutrino masses arise from $H$ and $\eta$. One solution for the minimum of the Higgs potential~\cite{arXiv:1007.0871} is 
\begin{equation}
  \vev{ H^0}=v_h\ne 0,~~~~ \vev{ \eta^0_1}=v_\eta \ne 0~~~~
\vev{ \eta^0_{2,3}}=0\,,
\end{equation}
which means the vev alignment for the $A_4$ triplet of the form $\vev{ \eta} \sim (1,0,0)$. This alignment is invariant under the $S$ generator\footnote{$H$ is in the $1$ representation of $A_4$ and its vev also respect the generator $S$.}, see eq. (\ref{generators}), which means that the minimum of the potential breaks spontaneously $A_4$ into a $Z_2$ subgroup generated by $S$. All the fields in the model singlets under $A_4$ are even under the residual $Z_2$, the triplets transform as:
\begin{equation}\label{residualZ2}
\begin{array}{lcrlcr}
N_1 &\to& +N_1\,,\quad& \eta_1 &\to& +\eta_1 \\   
N_2 &\to& -N_2\,,\quad& \eta_2 &\to& -\eta_2 \\   
N_3 &\to& -N_3\,,\quad& \eta_3 &\to& -\eta_3.  
\end{array}
\end{equation}
The DM candidate is the lightest particle charged under $Z_2$ i.e. the lightest combination of the scalars $\eta_2$ and $\eta_3$, which we will denote
generically by $\eta_{DM}$. We list below all interactions of $\eta_{DM}$:
\begin{enumerate}
\item Yukawa interactions
\begin{equation}
\begin{array}{l}
\eta_{_{DM}}\, \overline{\nu}_i   N_{2,3}\,,
\end{array}
\end{equation}
where $i=e,\,\mu,\,\tau$.
\item Higgs-Vector boson couplings
\begin{equation}\label{eq:gint}
\begin{array}{l}
\eta_{_{DM}}\eta_{_{DM}} ZZ\,,\quad \eta_{_{DM}}\eta_{_{DM}} WW\,,\quad
\eta_{_{DM}}\eta_{2,3}^{\pm} W^\pm Z\,,\quad \eta_{_{DM}}\eta_{2,3}^{\pm} W^\pm \,,\quad\eta_{_{DM}}A_{2,3} Z \,.
\end{array}
\end{equation}
\item Scalar interactions from the Higgs potential:
\begin{equation}\label{eq:Pint}
\begin{array}{l}
\eta_{_{DM}}\, A_1 A_2 h\,,\quad \eta_{_{DM}}\, A_1 A_3 h_1\,,\\
\eta_{_{DM}}\, A_1 A_2 h_1\,,\quad \eta_{_{DM}}\, A_1 A_3 h\,,\\ 
\eta_{_{DM}}\, A_2 A_3 h_3\,,\quad\eta_{_{DM}}\, h_1 h_3 h\,\\ 
\eta_{_{DM}}\eta_{_{DM}} hh\,,\quad\eta_{_{DM}}\eta_{_{DM}} h_1h_1\,.
\end{array}
\end{equation}
\end{enumerate}
After electroweak symmetry breaking, the vevs of the Higgs fields acquire vacuum expectation values, $v_h$ and $v_\eta$ for the singlet and the first component of the triplet respectively. Additional terms are obtained from those in Eq.~(\ref{eq:Pint}) by replacing $ h\to v_h$ and $h_1\to v_\eta$. These vertices are relevant for direct detection~\cite{arXiv:1101.2874}
 see Figure~\ref{diagram}. The phenomenology of dark matter of this model has been studied in detail in~\cite{arXiv:1101.2874}. The model accommodates WMAP and collider constraits and  is consistent with Xenon100 and CDMS on one hand and CoGent or DAMA on the other hand~\cite{arXiv:1101.2874}, see Fig.~\ref{diagram}. 
%

\section{Neutrino masses phenomenology}

The model contains four heavy right-handed neutrinos. It is a special
case, called (3,4), of the general type-I seesaw
mechanism\cite{Schechter:1980gr}.  After electroweak symmetry
breaking, it is characterized by Dirac and Majorana mass-matrix:
\begin{equation}
m_D=\left(
\begin{array}{cccc}
x_1&0&0&x_4\\
x_2&0&0&0\\
x_3&0&0&0\\
\end{array}
\right),\quad
 M_R=\left(
 \begin{array}{cccc}
 M_1&0&0&0\\
 0&M_1&0&0\\
 0&0&M_1&0\\
 0&0&0&M_2
 \end{array}
 \right).
\end{equation}
where $x_{1},x_2,x_3$ and $x_4$ are respectively proportional to
$y_1^\nu$, $y_2^\nu$, $y_3^\nu$ and $y_4^\nu$ of eq.\,(\ref{lag}) and
are of the order of the electroweak scale, while $M_{1,2}$ are assumed to be close to the unification scale.
Light neutrinos get Majorana masses by means of the type-I seesaw relation and the light-neutrinos mass matrix
has the form:
\begin{equation}
\label{mnu}
m_\nu=-m_{D_{3\times 4}}M_{R_{4\times 4}}^{-1}m_{D_{3\times 4}}^T\equiv
\left(
\begin{array}{ccc}
y^2& ab&ac\\
ab&b^2&bc\\
ac&bc&c^2
\end{array}
\right).
\end{equation}
%
This texture of the light neutrino mass matrix has a null eigenvalue
$m_3=0$ corresponding to the eigenvector $(0,\,-b/c,\,1)^T,$
\footnote{ Note that if we were to stick to the minimal
  (3,3)-type-I seesaw scheme, with just 3 SU(2) singlet states, one
  would find a projective nature of the effective tree-level light
  neutrino mass matrix with two zero eigenvalues, hence
  phenomenologically inconsistent. That is why we adopted the (3,4)
  scheme.}  implying a vanishing reactor mixing angle $ \theta_{13}=0$
and inverse hierarchy. 
The model implies a neutrinoless double beta decay effective mass parameter in the range 0.03 to 0.05~eV at
3~$\sigma$, within reach of upcoming experiments.

\section{Conclusions}
We have studied a model where the stability of the dark matter
particle arises from a flavor symmetry.  The $A_4$ non-abelian
discrete group accounts for both the observed pattern of neutrino
mixing and for DM stability.
We have analyzed the constraints that follow from electroweak
precision tests, collider searches and perturbativity.
We have also analyzed the prospects for direct and indirect dark
matter detection and found that, although the former already excludes
a large region in parameter space, we cannot constrain the mass of the
DM candidate.
In contrast, indirect DM detection is not yet sensitive enough to
probe our predictions. However, forecasted sensitivities indicate that
Fermi-LAT should start probing them in the near future.

All of the above relies mainly on the properties of the scalar sector
responsible for the breaking of the gauge and flavor symmetries. The motivation of our approach is to link the origin of dark matter  to
the origin of neutrino mass and the understanding of the pattern of
neutrino mixing, two of the most outstanding challenges in particle
physics today. At this level one may ask what are the possible tests
of this idea in the neutrino sector. 
We found an inverted neutrino mass hierarchy, hence a neutrinoless double
beta decay rate accessible to upcoming searches, while $\theta_{13}=0$ giving no CP violation in neutrino oscillations.
Note however that the connection of dark matter to neutrino properties
depends strongly on how the symmetry breaking sector couples to the leptons.

\section{Acknowledgments}
I thank my collaborators Sofiane Boucenna, Martin Hirsch, Stefano Morisi, Marco Taoso and Jose W. F. Valle for the work presented here. This work was supported by the Spanish MICINN under grants
FPA2008-00319/FPA, FPA2011-22975 and MULTIDARK CSD2009-00064
(Consolider-Ingenio 2010 Programme), by Prometeo/2009/091 (Generalitat
Valenciana), by the EU ITN UNILHC
PITN-GA-2009-237920 and by CONACyT (Mexico).


\section*{References}

\begin{thebibliography}{9}
\bibitem{arXiv:1007.0871}
  M.~Hirsch, S.~Morisi, E.~Peinado and J.~W.~F.~Valle,
  Phys.\ Rev.\ D\ {\bf 82} (2010) 116003
  [arXiv:1007.0871 [hep-ph]].
\bibitem{arXiv:1011.1371}
  D.~Meloni, S.~Morisi and E.~Peinado,
  Phys.\ Lett.\ B\ {\bf 697} (2011) 339
  [arXiv:1011.1371 [hep-ph]].
\bibitem{arXiv:1104.0178}
  D.~Meloni, S.~Morisi, E.~Peinado, S.~Morisi and E.~Peinado,
  Phys.\ Lett.\ B\ {\bf 703} (2011) 281
  [arXiv:1104.0178 [hep-ph]].
\bibitem{arXiv:1010.4963}
  Y.~Daikoku, H.~Okada and T.~Toma,
  arXiv:1010.4963 [hep-ph].

\bibitem{arXiv:1011.5753}
  Y.~Kajiyama and H.~Okada,
  Nucl.\ Phys.\ B\ {\bf 848} (2011) 303
  [arXiv:1011.5753 [hep-ph]].

\bibitem{arXiv:1111.4938}
  D.~A.~Eby and P.~H.~Frampton,
  arXiv:1111.4938 [hep-ph].
\bibitem{arXiv:1104.5676}
  R.~d.~A.~Toorop, F.~Bazzocchi and S.~Morisi,
  arXiv:1104.5676 [hep-ph].
\bibitem{arXiv:1101.2874}
  M.~S.~Boucenna, M.~Hirsch, S.~Morisi, E.~Peinado, M.~Taoso and J.~W.~F.~Valle,
  JHEP\ {\bf 1105} (2011) 037
  [arXiv:1101.2874 [hep-ph]].
\bibitem{Schechter:1980gr}
  J.~Schechter and J.~W.~F.~Valle,
  Phys.\ Rev.\ D {\bf 22} (1980) 2227.
\end{thebibliography}
\end{document}